\documentclass[two column, prA]{revtex4}%
\usepackage{amsmath}
\usepackage{graphicx}
\usepackage{amsfonts}
\usepackage{amssymb}
\usepackage{color}%
\providecommand{\U}[1]{\protect\rule{.1in}{.1in}}
\ifx\pdfoutput\relax\let\pdfoutput=\undefined\fi
\newcount\msipdfoutput
\ifx\pdfoutput\undefined\else
\ifcase\pdfoutput\else
\ifx\paperwidth\undefined\else
\ifdim\paperheight=0pt\relax\else\pdfpageheight\paperheight\fi
\ifdim\paperwidth=0pt\relax\else\pdfpagewidth\paperwidth\fi
\fi\fi\fi
\begin{document}
\title[ ]{Covariant photon current }
\author{Margaret Hawton}
\affiliation{Lakehead University, Thunder Bay, ON, Canada, P7B 5E1}
\email{mhawton@lakeheadu.ca}

\begin{abstract}
Based on the physical interpretation of the photon continuity equation derived
in [M. Hawton, Phys. Rev. A, \textbf{109}, 062221 (2024) ] the standard
Lagrangian is second quantized to obtain a Lorentz and gauge invariant theory
of single photons. The scalar potential is not independently second quantized
so all modes have positive definite norm. The continuity equation is
generalized by separating the material source current into a nonabsorbing term
describing propagation in a lossless transmission line and localizable single
photon emission and detection terms that do not require nonlocal separation of
transverse and longitudinal modes.

\end{abstract}
\maketitle

\section{Introduction}

In quantum field theory (QFT) particles are discrete excitations of classical
fields created by second quantized operators. In the electromagnetic (EM) case
these particles are called photons. Since photons are bosons, all whole
numbers $n$ of excitations can exist and a general state in photon Fock space
is an arbitrary linear combination of $n$-photon states. QFT is
charge-parity-time (CPT) invariant so\ photon coupling to charged matter is
described by a Hermitian EM potential operator that is an odd linear
combination of creation and annihilation terms \cite{GellMann}.

In classical EM the Maxwell equations (MEs) are derived from the Lorentz and
gauge invariant standard Lagrangian. There are problems in QFT, such as
renormalization and elimination of divergent quantities, for which it is
essential to deal only with with manifestly covariant equations \cite{CT,PS}.
Dirac quantized the EM field in 1927 \cite{Dirac} and, ideally, second
quantized equations should also be derived from a Lorentz and gauge invariant
Lagrangian. This goal has remained unrealized for close a century
\cite{Mathews}. 

Gupta \cite{Gupta} and Bleuler \cite{Bleuler} independently quantized all four
types of photons using Pauli's \cite{Pauli} indefinite metric. To achieve this
they added a term of the form $\left(  \partial_{\mu}A^{\mu}\right)  ^{2}$ to
the Lagrangian. A subsiduary condition is then required to regain the MEs that
describe physical states. This is the current text book approach to covariant
second quantization. States described by the vector potential,
$\widehat{\mathbf{A}}$, have positive norm but those described by the scalar
potential, $\widehat{\phi}$, have negative norm so an indefinite metric is
defined and physical states and the MEs are required to satisfy a subsidiary
condition. This second quantized theory is described in detail in \cite{CT}.\ 

Gauge-invariance of the split of total angular momentum into spin and orbital
parts has been a subject of intense debate and controversy
\cite{AMcontroversy}. The lack of a Lorentz and gauge invariant theory has
been a major obstacles to the resolution of this debate. Motivated by this
controversy and recent experiments on topological phases \cite{Topological},
optical scattering \cite{Scattering} and two dimentional materials, photonic
crystal waveguides and optical fibers \cite{SolidState}, Yang, Khosravi and
Jacob \cite{SpinOperator} derived a QED operator for photon spin.

A continuity equation for photon four-current density was derived in
\cite{Conserved}. In these expressions the scalar potential $\phi$ is not
independently second quantized; instead it contributes to the description of
the longitudinal photon current. Based on the physical interpretation of this
continuity equation we will show that the standard Lagrangian can be second
quantized directly to give a Lorentz and gauge independent theory. 

Physical one photon pulses coupled to transmission lines and optical circuits
are now routinely prepared in the laboratory \cite{Manufacturable}. An
important application of the theory derived in \cite{Conserved} and extended
here is to the emission of single photons by a material source and their
causal propagation in an optical circuit until they are annihilated in a
photon counting detector. Since only whole numbers of photons exist, creation
and annihilation require a probabilistic interpretation.

In the next Section we second quantize the standard Lagrangian to give the MEs
and Lorentz and gauge invariant expressions that do not require subsiduary
conditions. We will show that the four-current operator derived in
\cite{Conserved} describes propagation of a single photon in the Fock space of
any physical state and derive a localizable conserved photon four-current
density and an expression for single photon helicity. For application to
optical circuits the material source current will be separated into a
propagating nonabsorbing term and a localized term that describes single
photon emitters and photon counting detectors.

\section{Theory}

For completeness we repeat and extend the definitions in \cite{Conserved}: SI
units are used throughout. The contravariant space-time, wavevector and
momentum four-vectors are $x=x^{\mu}=\left(  ct,\mathbf{x}\right)  ,$
$k=\left(  \omega_{k}/c,\mathbf{k}\right)  $ and $p=\hbar k$ where
$h=2\pi\hbar$ is Planck's constant, $c$ is the speed of light in free space,
$kx=\omega_{k}t-\mathbf{k\cdot x}$ is invariant, the four-gradient is
$\partial=\left(  \partial_{ct},-\mathbf{\nabla}\right)  $, $\square
\equiv\partial_{ct}^{2}-\mathbf{\nabla}^{2}$, the four-potential is $A=\left(
\frac{\phi}{c},\mathbf{A}\right)  $, the photon current is $J_{p}=\left(
c\rho_{p},\mathbf{J}_{p}\right)  $ and the electric current is $J_{e}=\left(
c\rho_{e},\mathbf{J}_{e}\right)  =\left(  c\rho_{s},\partial_{t}%
\mathbf{P+\nabla}\times\mathbf{M+J}_{es}\right)  $ in a dielectric with
polarization $\mathbf{P}$, magnetization $\mathbf{M}$ and localized source and
sink four-currents $J_{es}$. The dressed photon current in a medium with
propagation speed $v=\left(  \varepsilon\mu\right)  ^{-1/2}$ will be called
$J_{pm}$. The covariant four-vector corresponding to $U^{\mu}=\left(
U_{0},\mathbf{U}\right)  $ is $U_{\mu}=g_{\mu\nu}U^{\nu}=\left(
U_{0},-\mathbf{U}\right)  $ where $g_{\mu\nu}=g^{\mu\nu}$ is a $4\times4$
diagonal matrix with diagonal $\left(  1,-1,-1,-1\right)  $ and $U_{\mu}%
U^{\mu}=U^{\mu}U_{\mu}$ is an invariant. The mutually orthogonal unit vectors
are $e^{\mu}$ where $e_{0}=n^{0}=\left(  1,0,0,0\right)  $ is time-like and,
in $\mathbf{k}$-space, $\mathbf{e}_{\mathbf{k}}=\mathbf{k}/\left\vert
\mathbf{k}\right\vert =\mathbf{e}_{\parallel}$ is longitudinal and the
definite helicity transverse unit vectors are $\mathbf{e}_{\lambda}\left(
\mathbf{k}\right)  =\frac{1}{\sqrt{2}}\left(  \mathbf{e}_{\theta}%
+i\lambda\mathbf{e}_{\phi}\right)  $ with $\lambda=\pm1$. If the propagation
direction is well defined to avoid confusion with $kx$ we choose wave vector
$\mathbf{k}=k\mathbf{e}_{z}$ and spatial coordinate $z$ so that $kx\rightarrow
-k\left(  z-vt\right)  $. The vector commutators will be written as $\left[
\widehat{\mathbf{V}}_{1},\cdot\widehat{\mathbf{V}}_{2}\right]
=\widehat{\mathbf{V}}_{1}\cdot\widehat{\mathbf{V}}_{2}-\widehat{\mathbf{V}%
}_{2}\cdot\widehat{\mathbf{V}}_{1}$ and $\left[  \widehat{\mathbf{V}}%
_{1},\times\widehat{\mathbf{V}}_{2}\right]  =\widehat{\mathbf{V}}_{1}%
\times\widehat{\mathbf{V}}_{2}-\widehat{\mathbf{V}}_{2}\times
\widehat{\mathbf{V}}_{1}$ for conciseness.

The inhomogeneous continuity equation%
\begin{equation}
\partial_{\mu}\widehat{J}_{p}^{\mu}=\partial_{t}\widehat{\rho}_{p}%
\mathbf{+\nabla}\cdot\widehat{\mathbf{J}}_{p}=\frac{-i\varepsilon_{0}c}%
{2\hbar}\left[  \widehat{A}_{\mu},\widehat{J}_{e}^{\mu}\right]
\label{Jcontinuity}%
\end{equation}
for four-current operator
\begin{align}
\widehat{J}_{p}  & =\left(  c\widehat{\rho}_{p},\widehat{\mathbf{J}}%
_{p}\right)  \label{fourcurrent}\\
& =\frac{-i\varepsilon_{0}c}{2\hbar}\left[  \widehat{A}_{\mu}%
,\widehat{\mathcal{F}}_{e}^{\mu\nu}\right]  \nonumber
\end{align}
was derived in \cite{Conserved}. The material source term $\frac
{-i\varepsilon_{0}c}{2\hbar}\left[  \widehat{A}_{\mu},\widehat{J}_{e}^{\mu
}\right]  $ describes polarization of the transmission line, single photon
emitters and photon counting detectors. The operator $\widehat{\rho}_{p}$ is
the norm of a one-photon state so%
\begin{equation}
\widehat{\rho}_{p}=\int d\mathbf{x}\rho_{p}\left(  x\right)
=1\label{normoperator}%
\end{equation}
where $\rho_{p}\left(  x\right)  $ is photon number density.where the Faraday
tensor
\begin{align}
\mathcal{F}^{\mu\nu}\left(  x\right)   &  =\partial^{\mu}A^{\nu}-\partial
^{\nu}A^{\mu}\label{F}\\
&  =\frac{1}{c}\left(
\begin{array}
[c]{cccc}%
0 & -E_{x} & -E_{y} & -E_{z}\\
E_{x} & 0 & -cB_{z} & cB_{y}\\
E_{y} & cB_{z} & 0 & -cB_{x}\\
E_{z} & -cB_{y} & cB_{x} & 0
\end{array}
\right)  \label{Faraday}%
\end{align}
is the four-dimensional curl of $A$. Substitution gives%
\begin{equation}
\widehat{J}_{p}=\frac{-i\varepsilon_{0}}{2\hbar}\left(  \left[
\widehat{\mathbf{A}},\cdot\widehat{\mathbf{E}}\right]  ,\left[
\widehat{\mathbf{A}},\times c\widehat{\mathbf{B}}\right]  +\left[
\frac{\widehat{\phi}}{c},\widehat{\mathbf{E}}\right]  \right)
.\label{Jexpanded}%
\end{equation}
Since $\mathbf{\nabla}\cdot\left(  \phi\mathbf{E}_{\perp}\right)
=\phi\mathbf{\nabla}\cdot\mathbf{E}_{\perp}+\left(  \mathbf{\nabla}%
\phi\right)  \cdot\mathbf{E}_{\perp}=0$ and $\mathbf{\nabla}\cdot\left(
\mathbf{A}_{\parallel}\times\mathbf{B}\right)  =\left(  \mathbf{\nabla}%
\times\mathbf{A}_{\parallel}\right)  \cdot\mathbf{B}-\mathbf{A}_{\parallel
}\cdot\left(  \mathbf{\nabla}\times\mathbf{B}\right)  =0$,%
\begin{equation}
\widehat{J}_{p}=\frac{-i\varepsilon_{0}}{2\hbar}\left(  \left[
\widehat{\mathbf{A}},\cdot\widehat{\mathbf{E}}\right]  ,\left[
\widehat{\mathbf{A}}_{\bot},\times c\widehat{\mathbf{B}}\right]  +\left[
\frac{\widehat{\phi}}{c},\widehat{\mathbf{E}}_{\parallel}\right]  \right)
\label{Jp}%
\end{equation}
which is separated into its transverse and longitudinal components is
equivalent to (\ref{Jexpanded}). Its spacetime components $\left[
\frac{\widehat{\phi}}{c},\widehat{\mathbf{E}}_{\parallel}\right]  $ extend
$\left[  \widehat{\mathbf{A}}_{\bot},\times c\widehat{\mathbf{B}}\right]  $ to
four-dimensions.The commutators in (\ref{Jcontinuity}) such as the norm
$\widehat{\rho}_{p}$ can be written as spatial integrals of densities.
Interchange of the order of differentiation and integration gives the
continuity equation%
\begin{equation}
\partial_{\mu}J_{p}^{\mu}\left(  x\right)  =\partial_{t}\rho_{p}\left(
x\right)  \mathbf{+\nabla}\cdot\mathbf{J}_{p}\left(  x\right)  =\frac
{-i\varepsilon_{0}c}{2\hbar}\left[  \widehat{A}_{\mu},\widehat{J}_{e}^{\mu
}\right]  .\label{continuity}%
\end{equation}

In (\ref{Jp}) the operators $\widehat{\mathbf{A}}$ and $\widehat{\mathbf{E}}$
are three vectors so, in general, $\widehat{\mathbf{A}}\cdot$
$\widehat{\mathbf{E}}$ is a sum over transverse and longitudinal polarizations
and the norm is positive definite. The scalar potential operator
$\widehat{\phi}$ is not separately second quantized, instead it appears in the
expression for the longitudinal current operator, $\widehat{\mathbf{J}}_{p}$,
so the indefinite scalar product is not required. In the next paragraph we
derive these covariant equations from the standard Lagrangian.

A Lagrangian completely defines the classical and second quantized equations.
Based on the MEs $\nabla\cdot\mathbf{B}\left(  x\right)  =0$ and $\nabla
\times\mathbf{E}\left(  x\right)  +\partial_{t}\mathbf{B}\left(  x\right)  =0$
the EM four-potential $\left(  \phi/c,\mathbf{A}\right)  $ can be defined such
that $\mathbf{B}\left(  x\right)  =\nabla\times\mathbf{A}\left(  x\right)  $
and $\mathbf{E}\left(  x\right)  =-\partial_{t}\mathbf{A}\left(  x\right)
-\nabla\phi\left(  x\right)  $. The invariant standard Lagrangian density is
then
\begin{equation}
\mathcal{L}=-\frac{1}{4}\varepsilon_{0}c^{2}\mathcal{F}_{\mu\nu}%
\mathcal{F}^{\mu\nu}-J_{e}^{\mu}A_{\mu}\label{Ldensity}%
\end{equation}
where $-\frac{1}{4}\varepsilon_{0}c^{2}\mathcal{F}_{\mu\nu}\mathcal{F}^{\mu
\nu}=\frac{1}{2}\varepsilon_{0}\left(  \mathbf{E}\cdot\mathbf{E}%
-c^{2}\mathbf{B}\cdot\mathbf{B}\right)  $ and $J_{e}$ is the electric
four-current. This is consistent with the quantum electrodynamic (QED)
Lagrangian $\mathcal{L}_{QED}=\mathcal{L}_{Dirac}+\mathcal{L}$ if $J_{e}$ is
the Dirac current \cite{PS}. The QED Lagrangian is invariant under the gauge
transformation%
\begin{equation}
A^{\mu}\rightarrow A^{\mu}-\frac{1}{e}\partial^{\mu}\alpha\left(  x\right)
\label{alpha}%
\end{equation}
where $e=-\left\vert e\right\vert $ is the charge on the electron. Since
$\square\alpha\left(  x\right)  =0$ this gauge transformation preserves the
invarance of $\partial_{\mu}A^{\mu}$. The covariant EM equations of motion are%
\begin{equation}
\varepsilon_{0}c^{2}\partial_{\mu}\mathcal{F}^{\mu\nu}=J_{e}^{\nu
}\label{EqMotion}%
\end{equation}
with scalar and vector components%
\begin{align}
\mathbf{\nabla}\cdot\mathbf{E} &  =\frac{\rho_{e}}{\varepsilon_{0}%
},\label{EM1}\\
\partial_{t}\mathbf{E}-c^{2}\mathbf{\nabla}\times\mathbf{B} &  \mathbf{=}%
-\frac{\mathbf{J}_{e}}{\varepsilon_{0}}.\label{EM2}%
\end{align}
Since $\mathbf{E=-\partial}_{t}\mathbf{A}-\mathbf{\nabla}\phi$ in
(\ref{Ldensity}) the momentum conjugate to $\mathbf{A}$ is
\begin{equation}
\mathbf{\Pi}=\partial\mathcal{L}/\partial\left(  \partial_{t}\mathbf{A}%
\right)  =-\varepsilon_{0}\mathbf{E.}\label{Pi}%
\end{equation}
Based on the usual rules for second quantization the commutation relation that
defines a one photon state is
\begin{equation}
\varepsilon_{0}\left[  \widehat{\mathbf{A}},\cdot\widehat{\mathbf{E}}\right]
=-i\hbar.\label{comm}%
\end{equation}
Any classical state with arbitrary polarization can be second quantized in
this way. The continuity equation equation for covariant single photon
four-current operator (\ref{Jp}) can be derived from the equations of motion
(\ref{EM1}) and (\ref{EM2}). Since $\widehat{\mathbf{E}}$ and
$\widehat{\mathbf{B}}$ commute with themselves, $\mathbf{\nabla}\cdot\left[
\left(  \widehat{\mathbf{A}}\mathbf{,}\times\widehat{\mathbf{B}}\right)
\right]  =-\left[  \widehat{\mathbf{A}}\cdot,\left(  \mathbf{\nabla}%
\times\widehat{\mathbf{B}}\right)  \right]  $ and $\left[  -\partial
_{t}\widehat{\mathbf{A}}-\mathbf{\nabla}\widehat{\phi},\cdot
\widehat{\mathbf{E}}\right]  =0$, substitution of these identities and
(\ref{Jp}) gives (\ref{Jcontinuity}), verifying the photon continuity equation.

According to the MEs (\ref{EM1}) and (\ref{EM2}) the EM field in a charge free
region is transverse while a localized charge gives rise to a longitudinal
electric field, $\mathbf{E}_{\parallel}$. The field $\mathbf{E}_{\parallel}$
is outward if the charge is positive and inward if it is negative. By
inspection of (\ref{Jp}) it follows that the photon number current is also
outward if the electric charge is positive and inward if it is negative. For a
stationary charge the inhomogeneous continuity equaton (\ref{Jcontinuity})
reduces to $\partial_{t}\widehat{\rho}_{p}\mathbf{+\nabla}\cdot
\widehat{\mathbf{J}}_{p}=\frac{-i\varepsilon_{0}}{2\hbar}\left[
\widehat{\phi},\widehat{\rho}_{e}\right]  $.

In QED an arbitrary classical field mode is treated as a collection of
harmonic oscillators with definite frequency, wavevector and polarization.
Following \cite{SZ} as in \cite{Conserved}, in the discrete plane wave basis
the $n$-photon commutators, annihilation operators, creation operators and
expectation values for $n_{\mathbf{k}\lambda}$-photon states for transverse
and longitudinal modes $\lambda=\pm1,\parallel$ are%
\begin{align}
\left[  \widehat{a}_{\mathbf{k}\lambda},\widehat{a}_{\mathbf{k}^{\prime
}\lambda^{\prime}}\right]   &  =0,\ \left[  \widehat{a}_{\mathbf{k}\lambda
}^{\dag},\widehat{a}_{\mathbf{k}^{\prime}\lambda^{\prime}}^{\dagger}\right]
=0\label{commute}\\
\left[  \widehat{a}_{\mathbf{k}\lambda},\widehat{a}_{\mathbf{k}^{\prime
}\lambda^{\prime}}^{\dagger}\right]   &  =\widehat{a}_{\mathbf{k}\lambda
}\widehat{a}_{\mathbf{k}^{\prime}\lambda^{\prime}}^{\dagger}-\widehat{a}%
_{\mathbf{k}^{\prime}\lambda^{\prime}}^{\dagger}\widehat{a}_{\mathbf{k}%
\lambda}\nonumber\\
&  =\delta_{\lambda\lambda^{\prime}}\delta_{\mathbf{kk}^{\prime}%
}\label{dontcommute}\\
\widehat{a}_{\mathbf{k}\lambda n} &  \equiv\frac{\left(  \widehat{a}%
_{\mathbf{k}\lambda}\right)  ^{n}}{\sqrt{n!}},\ \widehat{a}_{\mathbf{k}\lambda
n}^{\dagger}=\left(  \widehat{a}_{\mathbf{k}\lambda n}\right)  ^{\dagger
},\label{nphotons}\\
\ \left\vert n_{\mathbf{k}\lambda}\right\rangle  &  =\widehat{a}%
_{\mathbf{k}\lambda n}\left\vert 0\right\rangle ,\label{nphotonstate}\\
\left\langle n_{\mathbf{k}\lambda}\left\vert \widehat{a}_{\mathbf{k}\lambda
}^{\dag}\widehat{a}_{\mathbf{k}\lambda}\right\vert n_{\mathbf{k}\lambda
}\right\rangle  &  =n_{\mathbf{k}\lambda},\ \\
\left\langle n_{\mathbf{k}\lambda}\left\vert \widehat{a}_{\mathbf{k}\lambda
}\widehat{a}_{\mathbf{k}\lambda}^{\dag}\right\vert n_{\mathbf{k}\lambda
}\right\rangle  &  =\left\langle n_{\mathbf{k}\lambda}+1|n_{\mathbf{k}\lambda
}+1\right\rangle =n_{\mathbf{k}\lambda}+1
\end{align}%
\begin{equation}
\left\langle n_{\mathbf{k}\lambda}\left\vert \widehat{a}_{\mathbf{k}\lambda
}\widehat{a}_{\mathbf{k}\lambda}^{\dagger}-\widehat{a}_{\mathbf{k}\lambda
}^{\dagger}\widehat{a}_{\mathbf{k}\lambda}\right\vert n_{\mathbf{k}\lambda
}\right\rangle =n_{\mathbf{k}\lambda}+1-n_{\mathbf{k}\lambda}%
=1.\label{countsone}%
\end{equation}
Eq. (\ref{countsone}) implies that the expectation value of the commutator and
hence the current density operator, (\ref{Jp}), does not depend on $\left\vert
n_{\mathbf{k}\lambda}\right\rangle $ and describes the addition of one photon
to any Fock state. In the continuum limit $\Delta n/V\rightarrow
d\mathbf{k}/\left(  2\pi\right)  ^{3}$ and $\int d\mathbf{k/2\omega}_{k}%
(2\pi)^{3}$ is an invariant so we define the plane wave basis
\begin{equation}
\left[  \widehat{a}_{\lambda}\left(  \mathbf{k}\right)  ,\widehat{a}%
_{\lambda^{\prime}}^{\dagger}\left(  \mathbf{k}^{\prime}\right)  \right]
=\delta_{\lambda\lambda^{\prime}}2\omega_{k}\delta\left(  \mathbf{k-k}%
^{\prime}\right)  .\label{kcommutation}%
\end{equation}
The transverse unit vectors%
\[
\mathbf{e}_{\lambda}\left(  \mathbf{k}\right)  =\frac{1}{\sqrt{2}}\left(
\mathbf{e}_{\theta}+i\lambda\mathbf{e}_{\phi}\right)
\]
satisfy the orthonormality relations%
\begin{align}
\mathbf{e}_{\lambda}^{\ast}\cdot\mathbf{e}_{\lambda^{\prime}}  &
=\delta_{\lambda\lambda^{\prime}},\label{edote}\\
\mathbf{e}_{\lambda}^{\ast}\times\mathbf{e}_{\lambda^{\prime}}  &
=i\lambda\delta_{\lambda\lambda^{\prime}}\mathbf{e}_{\mathbf{k}}%
\label{ecrosse}%
\end{align}
The covariant vector potential, electric and \ magnetic field operators are
then \
\begin{align}
\widehat{\mathbf{A}}^{+}\left(  x\right)   &  =i\sqrt{\frac{\hbar
}{2\varepsilon_{0}}}\sum_{\lambda=\pm1,\parallel}\int\frac{d\mathbf{k}%
}{\left(  2\pi\right)  ^{3}2\omega_{k}}\nonumber\\
&  \times\widehat{a}_{\lambda}\left(  \mathbf{k}\right)  c_{\lambda}\left(
\mathbf{k}\right)  \mathbf{e}_{\lambda}\left(  \mathbf{k}\right)
e^{-ikx},\label{Apluslambdanop}\\
\widehat{\mathbf{A}}^{-} &  =\widehat{\mathbf{A}}^{+\dagger}%
,\ \widehat{\mathbf{A}}=\widehat{\mathbf{A}}^{+}+\widehat{\mathbf{A}}%
^{-},\ \label{Aminusandtotal}\\
\ \widehat{\mathbf{E}} &  =-\partial_{t}\widehat{\mathbf{A}}-\mathbf{\nabla
}\widehat{\phi},\ \ \widehat{\mathbf{B}}=\mathbf{\nabla}\times
\widehat{\mathbf{A}},\label{EandB}\\
\text{and }\mathbf{e}_{\lambda}\left(  \mathbf{k}\right)   &  \rightarrow
\mathbf{\lambda}k\mathbf{e}_{\lambda}\left(  \mathbf{k}\right)  \text{ in
}\widehat{\mathbf{B}}_{\lambda}\text{.}\label{Blambda}%
\end{align}
The superscript $\dagger$ is the Hermitian conjugate, $\pm$ refer to positive
and negative frequency parts and $c_{\lambda}\left(  \mathbf{k}\right)  $ is
the invariant probability amplitude for wave vector $\mathbf{k}$ and
polarization $\lambda=\pm1,\parallel$. Substitution in (\ref{comm}) gives%
\begin{equation}
\frac{i\varepsilon_{0}}{\hbar}\left[  \widehat{\mathbf{A}},\widehat{\cdot
\mathbf{E}}\right]  =\sum_{\lambda=\pm1,\parallel}\int\frac{d\mathbf{k}%
}{\left(  2\pi\right)  ^{3}2\omega_{k}}c_{\lambda}^{\ast}\left(
\mathbf{k}\right)  c_{\lambda}\left(  \mathbf{k}\right)  =1\label{kspacenorm}%
\end{equation}
which is the norm of a one photon state in $\mathbf{k}$-space. The spatial
integral of the photon four-current (\ref{Jp}) evaluated in $\mathbf{k}$-space
is%
\begin{equation}
\int d\mathbf{x}J_{p}\left(  x\right)  =\sum_{\lambda=\pm1,\parallel}\int%
\frac{d\mathbf{k}}{\left(  2\pi\right)  ^{3}}c_{\lambda}^{\ast}\left(
\mathbf{k}\right)  c_{\lambda}\left(  \mathbf{k}\right)  \left(
1,\mathbf{e}_{\mathbf{k}}\right)  .\label{Jinkspace}%
\end{equation}
In any gauge that satisfies $\partial_{\mu}A^{\mu}=c^{-2}\partial_{t}%
\phi+\mathbf{\nabla}\cdot\mathbf{A}=0$ \
\begin{equation}
\phi\left(  x\right)  =cA_{\parallel}\left(  x\right)  .\label{phi}%
\end{equation}
Eqs. (\ref{Apluslambdanop}) to (\ref{Jinkspace}) and (\ref{Jcontinuity}) to
(\ref{Jp}) are valid for any normalizable sum over wavevectors $\mathbf{k}$
and polarizations $\lambda=\pm1,\parallel$ described by $\left\{  c_{\lambda
}\left(  \mathbf{k}\right)  \right\}  $. Any classical electromagnetic mode
can be second quantized in this way.

According to (\ref{commute}) creation and annihilation operators commute
amongst themselves so the free space four-current operator describing creation
of one photon can be written as%
\begin{align}
\int d\mathbf{x}J_{p}\left(  x\right)   &  =-\frac{i\varepsilon_{0}}{2\hbar
}\left(  \left[  \widehat{\mathbf{A}}^{+},\cdot\widehat{\mathbf{E}}%
^{-}\right]  ,-\left[  \widehat{\mathbf{A}}^{+},\times c\widehat{\mathbf{B}%
}^{-}\right]  \right.  \nonumber\\
&  \left.  +\left[  \frac{\widehat{\phi}^{+}}{c},\widehat{\mathbf{E}%
}_{\parallel}^{-}\right]  +H.c.\right)  \label{Jopxspace}%
\end{align}
where $H.c.$ is the Hermitian conjugate. Since $\left[  \widehat{\mathbf{A}%
},\cdot\widehat{\mathbf{E}}\right]  =\left[  \widehat{\mathbf{A}}^{+}%
,\cdot\widehat{\mathbf{E}}^{-}\right]  +H.c.$,
\begin{equation}
\rho_{p}\left(  x\right)  =-\frac{i\varepsilon_{0}}{2\hbar}\mathbf{A}%
^{+}\left(  x\right)  \cdot\mathbf{E}^{-}\left(  x\right)  +c.c.\label{rho}%
\end{equation}
where $c.c.$ is the complex conjugate so the number density, $\rho_{p}\left(
x\right)  $, is real. Only whole numbers of photons exist so the normalization
$\widehat{\rho}_{p}=1$ is preserved until the whole photon is annihilated. 

Symmetries that lead to conservation laws for angular momenta are determined
by a particle's Wigner little group \cite{Weinberg}. Massive particles have a
rest frame so their spherically symmetrical little group consists of rotations
in three dimensions. The photon little group is cylindrically symmetrical and
includes an operator that generates rotations about some fixed but arbitrary
axis. A realization of the photon little group based on the photon position
operators is described in \cite{HawtonDebierre}. Based on (\ref{edote}) and
(\ref{ecrosse}) the photon helicity density is%
\begin{equation}
\mathbf{S}_{\lambda}\left(  x\right)  =-\frac{i\varepsilon_{0}}{2}%
\mathbf{A}_{\lambda}^{+}\left(  x\right)  \times\mathbf{E}_{\lambda}%
^{-}\left(  x\right)  +c.c.=\lambda\hbar\rho_{p}\left(  x\right)
\mathbf{e}_{\mathbf{k}}.\label{helicity}%
\end{equation}
This is similar to the expression derived in \cite{SpinOperator} except that
$\mathbf{A}\cdot\mathbf{E}$ is replaced with $\mathbf{\operatorname{Re}%
}\left(  \mathbf{A}^{+}\cdot\mathbf{E}^{-}\right)  \ $here, allowing for use
of the complex form of the transverse unit vectors. For longitudinal modes
(\ref{helicity}) gives $0$.

In the continuity equation (\ref{Jcontinuity}) polarization and magnetization
of the medium act as external driving forces. However, many recent experiments
involve propagation in transmission lines and optical circuits. At infrared
and visible frequencies a medium can be treated as continuous by averaging
over domains of order $10^{-8}m$ \cite{Conserved}. In the presence of
localized sources and sinks the electric current operator can be written as%
\begin{align}
\widehat{\mathbf{J}}_{e} &  =\partial_{t}\widehat{\mathbf{P}}+\mathbf{\nabla
}\times\widehat{\mathbf{M}}+\widehat{\mathbf{J}}_{es},\label{PandM}\\
\widehat{J}_{es} &  =\left(  c\widehat{\rho}_{es},\ \widehat{\mathbf{J}}%
_{es}\right)  .\label{barerho}%
\end{align}
where $\mathbf{P}$ is polarization, $\mathbf{M}$ is magnetization and electric
displacement and magnetic field operators are
\begin{align}
\widehat{\mathbf{D}} &  =\varepsilon_{0}\widehat{\mathbf{E}}%
+\widehat{\mathbf{P}}=\varepsilon\widehat{\mathbf{E}},\label{D}\\
\widehat{\mathbf{H}} &  =\mu_{0}^{-1}\widehat{\mathbf{B}}-\widehat{\mathbf{M}%
}=\mu^{-1}\widehat{\mathbf{B}}.\label{H}%
\end{align}
The norm of a one photon state,%
\begin{equation}
\frac{-i\varepsilon_{0}}{2\hbar c}\left[  \widehat{\mathbf{A}},\cdot
\widehat{\mathbf{E}}\right]  =\int d\mathbf{x}\rho_{p}\left(  x\right)
=1,\label{norm}%
\end{equation}
that is material independent will be retained in a polarizable medium since
replacement of $\varepsilon_{0}$ with $\varepsilon$ includes polarization density.

Substitution of (\ref{PandM}) to (\ref{H}) in the continuity equation
(\ref{Jcontinuity}) gives the four-current operator in a medium,%
\begin{equation}
\widehat{J}_{pm}\left(  x\right)  =\frac{-i}{2\hbar}\left(  \left[
\widehat{\mathbf{A}},\cdot\widehat{\mathbf{D}}\right]  ,-c\varepsilon
\mu\left[  \widehat{\mathbf{A}}_{\perp},\times\widehat{\mathbf{H}}\right]
+\left[  \frac{\widehat{\phi}}{c},\widehat{\mathbf{D}}_{\parallel}\right]
\right)  .\label{Jm}%
\end{equation}
The localized part of the invariant source term in (\ref{continuity}) that
describes single photon emitters and photon counting detectors can be written
as%
\begin{equation}
\frac{-i\varepsilon_{0}c}{2\hbar}\left[  \widehat{A}_{\mu},\widehat{J}%
_{es}^{\mu}\right]  =\int d\mathbf{x}\left\{  \partial_{t}\rho_{es}\left(
x\right)  +\mathbf{\nabla}\cdot\mathbf{J}_{es}\left(  x\right)  \right\}
\label{sourceop}%
\end{equation}
so the position space inhomgeneous continuity equation in a medium in the
presence of localized emitters and detectors is%
\begin{equation}
\partial_{t}\rho_{pm}\left(  x\right)  \mathbf{+\nabla}\cdot\mathbf{J}%
_{pm}\left(  x\right)  =\partial_{t}\rho_{es}\left(  x\right)  +\mathbf{\nabla
}\cdot\mathbf{J}_{es}\left(  x\right)  .\label{withsource}%
\end{equation}
This is the general case since it describes propagation in free space if
$\varepsilon=\varepsilon_{0}$ and $\mu=\mu_{0}$. Since $\int dt\int
d\mathbf{x}\delta\left(  t-t^{\prime}\right)  \delta\left(  \mathbf{x}%
-\mathbf{x}^{\prime}\right)  =1$, $\delta\left(  t-t^{\prime}\right)
\delta\left(  \mathbf{x}-\mathbf{x}^{\prime}\right)  $ describes an
instantaneous localized source, the equation for the Green's function operator
$J_{px^{\prime}}$ for the response to this instantaneous localized source at
time $t^{\prime}$and position $\mathbf{x}^{\prime}$ is%
\begin{equation}
\partial_{t}\rho_{px^{\prime}}\mathbf{+\nabla}\cdot\mathbf{J}_{px^{\prime}%
}=\delta\left(  t-t^{\prime}\right)  \delta\left(  \mathbf{x}-\mathbf{x}%
^{\prime}\right)  .\label{Green}%
\end{equation}
The equation describing creation, propagation and detection of a single photon
is then%
\begin{align}
\partial_{t}\rho_{pm}\mathbf{+\nabla}\cdot\mathbf{J}_{pm} &  =\int dt^{\prime
}\int d\mathbf{x}^{\prime}\left\{  \partial_{t}\rho_{es}\left(  x-x^{\prime
}\right)  \right.  \nonumber\\
&  \left.  +\mathbf{\nabla}\cdot\mathbf{J}_{es}\left(  x-x^{\prime}\right)
\right\}  .\label{solution}%
\end{align}
The real localizable photon number density,
\begin{equation}
\rho_{p}\left(  x\right)  =\frac{\varepsilon_{0}}{\varepsilon}\rho_{pm}\left(
x\right)  ,\label{photonnodensity}%
\end{equation}
is determined by the position and time of the photon's creation and its
direction of propagation.

A one dimensional approximation provides a useful description of propagation
of a beam in free space or a dielectric transmission line. For $z$-axis chosen
parallel to the direction of propagation of a beam with uniform
cross-sectional area and direction of propagation $\mathbf{e}_{z}$ the
four-current density%
\begin{align*}
J_{pm} &  =\left(  v\rho_{pm},\mathbf{J}_{pm}\right)  ,\\
\mathbf{J}_{pm} &  =v\rho_{pm}\mathbf{e}_{\mathbf{z}}%
\end{align*}
satisfies a continuity equation.

\section{Conclusion}

At a fundamental level the EM field strength is not continuous, it is a Fock
space of $n$-photon states. Physical states are normalizable and the
continuity equations (\ref{Jcontinuity}) and (\ref{solution}) that describe
the conservation of photon number for the four-current operator (\ref{Jm}) are
valid for any physical state. The commutator describes emission, propagation
and detection of a single photon with norm one. Consistent with classical and
second quantized electromagnetism, this photon does not interact with any
other photons present in Fock space.

It is not generally accepted that a photon is localizable \cite{DeBievre} and
the energy density of a single photon state has been proved to be nonlocal
\cite{Federicoetal} but a photon emitted by a localized source \cite{Beige}
propagates causally so that at a later time it is no farther from the source
than a distance $vt$ where $v\leq c.$ This requires localization in a bounded
region of space. The continuity equation is valid in any gauge, but the
Coulomb gauge in which $\mathbf{A}_{\parallel}=0$ requires the non-local
separation of the longitudinal and transverse modes in the emitters and
detectors and this only complicates the calculation. The photon number density
(\ref{photonnodensity}) is real and it propagates causally. It is localizable
because it reflects the position and time that the photon was emitted by a
localized material source.

The transverse modes propagate in free space or in a transmission line and
their discreteness is made macroscopically observable when a photon is counted
by reducing an $n$-photon state to an $\left(  n-1\right)  $-photon state. A
simple example of this is experimentally verified in \cite{BeamSplitter} in
which a single photon is injected into an optical circuit consisting of a
biprism that splits the photon density into two paths, each terminated with a
photon counting detector. Within experimental error the photon was counted in
only one of these detectors. This signifies nonlocal collapse in the photon
Fock space.

Here we derived these covariant second quantized equations from the standard
EM Lagrangian (\ref{Ldensity}). If the electric current is the Dirac current
(\ref{Ldensity}) is consistent with the QED Lagrangian. The scalar potential
is not independently second quantized, so the norm of any single photon state
is positive definite and equal to unity.

\end{document}